\pdfoutput=1
\documentclass[12pt,a4paper,titlepage]{article}
\usepackage[utf8]{inputenc}
\usepackage[top=50pt,bottom=50pt,left=68pt,right=66pt]{geometry}
\usepackage{amsmath}
\usepackage{booktabs}
\usepackage{amssymb}
\usepackage[title]{appendix}
\usepackage{mathrsfs}
\usepackage{float}
\usepackage{multirow}
\usepackage{graphicx,caption,subcaption}
\usepackage[space]{grffile}

\begin{document}
\renewcommand{\arraystretch}{1.3}

\makeatletter
\def\@hangfrom#1{\setbox\@tempboxa\hbox{{#1}}%
      \hangindent 0pt%\wd\@tempboxa
      \noindent\box\@tempboxa}
\makeatother

% Underline for text or math

\def\un#1{\relax\ifmmode\@@underline#1\else
        $\@@underline{\hbox{#1}}$\relax\fi}

% Accents and foreign (in text):

\let\under=\unt                 % bar-under (but see \un above)
\let\ced=\ce                    % cedilla
\let\du=\du                     % dot-under
\let\um=\Hu                     % Hungarian umlaut
\let\sll=\lp                    % slashed (suppressed) l (Polish)
\let\Sll=\Lp                    % " L
\let\slo=\os                    % slashed o (Scandinavian)
\let\Slo=\Os                    % " O
\let\tie=\ta                    % tie-after (semicircle connecting two letters)
\let\br=\ub                     % breve
                % Also: \`        grave
                %       \'        acute
                %       \v        hacek (check)
                %       \^        circumflex (hat)
                %       \~        tilde (squiggle)
                %       \=        macron (bar-over)
                %       \.        dot (over)
                %       \"        umlaut (dieresis)
                %       \aa \AA   A-with-circle (Scandinavian)
                %       \ae \AE   ligature (Latin & Scandinavian)
                %       \oe \OE   " (French)
                %       \ss       es-zet (German sharp s)
                %       \$  \#  \&  \%  \pounds  {\it\&}  \dots

% Abbreviations for Greek letters

\def\a{\alpha}
\def\b{\beta}
\def\c{\chi}
\def\d{\delta}
\def\e{\epsilon}
\def\f{\phi}
\def\g{\gamma}
\def\h{\eta}
\def\i{\iota}
\def\j{\psi}
\def\k{\kappa}
\def\l{\lambda}
\def\m{\mu}
\def\n{\nu}
\def\o{\omega}
\def\p{\pi}
\def\q{\theta}
\def\r{\rho}
\def\s{\sigma}
\def\t{\tau}
\def\u{\upsilon}
\def\x{\xi}
\def\z{\zeta}
\def\D{\Delta}
\def\F{\Phi}
\def\G{\Gamma}
\def\J{\Psi}
\def\L{\Lambda}
\def\O{\Omega}
\def\P{\Pi}
\def\Q{\Theta}
\def\S{\Sigma}
\def\U{\Upsilon}
\def\X{\Xi}

% Varletters

\def\ve{\varepsilon}
\def\vf{\varphi}
\def\vr{\varrho}
\def\vs{\varsigma}
\def\vq{\vartheta}

% Calligraphic letters

\def\ca{{\cal A}}
\def\cb{{\cal B}}
\def\cc{{\cal C}}
\def\cd{{\cal D}}
\def\ce{{\cal E}}
\def\cf{{\cal F}}
\def\cg{{\cal G}}
\def\ch{{\cal H}}
\def\ci{{\cal I}}
\def\cj{{\cal J}}
\def\ck{{\cal K}}
\def\cl{{\cal L}}
\def\cm{{\cal M}}
\def\cn{{\cal N}}
\def\co{{\cal O}}
\def\cp{{\cal P}}
\def\cq{{\cal Q}}
\def\car{{\cal R}}
\def\cs{{\cal S}}
\def\ct{{\cal T}}
\def\cu{{\cal U}}
\def\cv{{\cal V}}
\def\cw{{\cal W}}
\def\cx{{\cal X}}
\def\cy{{\cal Y}}
\def\cz{{\cal Z}}

% Fonts

\def\Sc#1{{\hbox{\sc #1}}}      % script for single characters in equations
\def\Sf#1{{\hbox{\sf #1}}}      % sans serif for single characters in equations

                        % Also:  \rm      Roman (default for text)
                        %        \bf      boldface
                        %        \it      italic
                        %        \mit     math italic (default for equations)
                        %        \sl      slanted
                        %        \em      emphatic
                        %        \tt      typewriter
                        % and sizes:    \tiny
                        %               \scriptsize
                        %               \footnotesize
                        %               \small
                        %               \normalsize
                        %               \large
                        %               \Large
                        %               \LARGE
                        %               \huge
                        %               \Huge

% Math symbols

\def\slpa{\slash{\pa}}                            % slashed partial derivative
\def\slin{\SLLash{\in}}                                   % slashed in-sign
\def\bo{{\raise-.3ex\hbox{\large$\Box$}}}               % D'Alembertian
\def\cbo{\Sc [}                                         % curly "
\def\pa{\partial}                                       % curly d
\def\de{\nabla}                                         % del
\def\dell{\bigtriangledown}                             % hi ho the dairy-o
\def\su{\sum}                                           % summation
\def\pr{\prod}                                          % product
\def\iff{\leftrightarrow}                               % <-->
\def\conj{{\hbox{\large *}}}                            % complex conjugate
\def\ltap{\raisebox{-.4ex}{\rlap{$\sim$}} \raisebox{.4ex}{$<$}}   % < or ~
\def\gtap{\raisebox{-.4ex}{\rlap{$\sim$}} \raisebox{.4ex}{$>$}}   % > or ~
\def\TH{{\raise.2ex\hbox{$\displaystyle \bigodot$}\mskip-4.7mu \llap H \;}}
\def\face{{\raise.2ex\hbox{$\displaystyle \bigodot$}\mskip-2.2mu \llap {$\ddot
        \smile$}}}                                      % happy face
\def\dg{\sp\dagger}                                     % hermitian conjugate
\def\ddg{\sp\ddagger}                                   % double dagger
                        % Also:  \int  \oint              integral, contour
                        %        \hbar                    h bar
                        %        \infty                   infinity
                        %        \sqrt                    square root
                        %        \pm  \mp                 plus or minus
                        %        \cdot  \cdots            centered dot(s)
                        %        \oplus  \otimes          group theory
                        %        \equiv                   equivalence
                        %        \sim                     ~
                        %        \approx                  approximately =
                        %        \propto                  funny alpha
                        %        \ne                      not =
                        %        \le \ge                  < or = , > or =
                        %        \{  \}                   braces
                        %        \to  \gets               -> , <-
                        % and spaces:  \,  \:  \;  \quad  \qquad
                        %              \!                 (negative)

\font\tenex=cmex10 scaled 1200

% Math stuff with one argument

\def\sp#1{{}^{#1}}                              % superscript (unaligned)
\def\sb#1{{}_{#1}}                              % sub"
\def\oldsl#1{\rlap/#1}                          % poor slash
\def\slash#1{\rlap{\hbox{$\mskip 1 mu /$}}#1}      % good slash for lower case
\def\Slash#1{\rlap{\hbox{$\mskip 3 mu /$}}#1}      % " upper
\def\SLash#1{\rlap{\hbox{$\mskip 4.5 mu /$}}#1}    % " fat stuff (e.g., M)
\def\SLLash#1{\rlap{\hbox{$\mskip 6 mu /$}}#1}      % slash for no-in sign
\def\PMMM#1{\rlap{\hbox{$\mskip 2 mu | $}}#1}   %
\def\PMM#1{\rlap{\hbox{$\mskip 4 mu ~ \mid $}}#1}       %
\def\Tilde#1{\widetilde{#1}}                    % big tilde
\def\Hat#1{\widehat{#1}}                        % big hat
\def\Bar#1{\overline{#1}}                       % big bar
\def\sbar#1{\stackrel{*}{\Bar{#1}}}             % big bar with star
\def\bra#1{\left\langle #1\right|}              % < |
\def\ket#1{\left| #1\right\rangle}              % | >
\def\VEV#1{\left\langle #1\right\rangle}        % < >
\def\abs#1{\left| #1\right|}                    % | |
\def\leftrightarrowfill{$\mathsurround=0pt \mathord\leftarrow \mkern-6mu
        \cleaders\hbox{$\mkern-2mu \mathord- \mkern-2mu$}\hfill
        \mkern-6mu \mathord\rightarrow$}
\def\dvec#1{\vbox{\ialign{##\crcr
        \leftrightarrowfill\crcr\noalign{\kern-1pt\nointerlineskip}
        $\hfil\displaystyle{#1}\hfil$\crcr}}}           % <--> accent
\def\dt#1{{\buildrel {\hbox{\LARGE .}} \over {#1}}}     % dot-over for sp/sb
\def\dtt#1{{\buildrel \bullet \over {#1}}}              % alternate "
\def\der#1{{\pa \over \pa {#1}}}                % partial derivative
\def\fder#1{{\d \over \d {#1}}}                 % functional derivative
                % Also math accents:    \bar
                %                       \check
                %                       \hat
                %                       \tilde
                %                       \acute
                %                       \grave
                %                       \breve
                %                       \dot    (over)
                %                       \ddot   (umlaut)
                %                       \vec    (vector)

% Math stuff with more than one argument

\def\frac#1#2{{\textstyle{#1\over\vphantom2\smash{\raise.20ex
        \hbox{$\scriptstyle{#2}$}}}}}                   % fraction
\def\half{\frac12}                                        % 1/2
\def\sfrac#1#2{{\vphantom1\smash{\lower.5ex\hbox{\small$#1$}}\over
        \vphantom1\smash{\raise.4ex\hbox{\small$#2$}}}} % alternate fraction
\def\bfrac#1#2{{\vphantom1\smash{\lower.5ex\hbox{$#1$}}\over
        \vphantom1\smash{\raise.3ex\hbox{$#2$}}}}       % "
\def\afrac#1#2{{\vphantom1\smash{\lower.5ex\hbox{$#1$}}\over#2}}    % "
\def\partder#1#2{{\partial #1\over\partial #2}}   % partial derivative of
\def\parvar#1#2{{\d #1\over \d #2}}               % variation of
\def\secder#1#2#3{{\partial^2 #1\over\partial #2 \partial #3}}  % second "
\def\on#1#2{\mathop{\null#2}\limits^{#1}}               % arbitrary accent
\def\bvec#1{\on\leftarrow{#1}}                  % backward vector accent
\def\oover#1{\on\circ{#1}}                              % circle accent

\def\[{\lfloor{\hskip 0.35pt}\!\!\!\lceil}
\def\]{\rfloor{\hskip 0.35pt}\!\!\!\rceil}
\def\Lag{{\cal L}}
\def\du#1#2{_{#1}{}^{#2}}
\def\ud#1#2{^{#1}{}_{#2}}
\def\dud#1#2#3{_{#1}{}^{#2}{}_{#3}}
\def\udu#1#2#3{^{#1}{}_{#2}{}^{#3}}
\def\calD{{\cal D}}
\def\calM{{\cal M}}

\def\szet{{${\scriptstyle \b}$}}
\def\ulA{{\un A}}
\def\ulM{{\underline M}}
\def\cdm{{\Sc D}_{--}}
\def\cdp{{\Sc D}_{++}}
\def\vTheta{\check\Theta}
\def\fracm#1#2{\hbox{\large{${\frac{{#1}}{{#2}}}$}}}
\def\ha{{\fracmm12}}
\def\tr{{\rm tr}}
\def\Tr{{\rm Tr}}
\def\itrema{$\ddot{\scriptstyle 1}$}
\def\ula{{\underline a}} \def\ulb{{\underline b}} \def\ulc{{\underline c}}
\def\uld{{\underline d}} \def\ule{{\underline e}} \def\ulf{{\underline f}}
\def\ulg{{\underline g}}
\def\items#1{\\ \item{[#1]}}
\def\ul{\underline}
\def\un{\underline}
\def\fracmm#1#2{{{#1}\over{#2}}}
\def\footnotew#1{\footnote{\hsize=6.5in {#1}}}
\def\low#1{{\raise -3pt\hbox{${\hskip 0.75pt}\!_{#1}$}}}

\def\Dot#1{\buildrel{_{_{\hskip 0.01in}\bullet}}\over{#1}}
\def\dt#1{\Dot{#1}}

\def\DDot#1{\buildrel{_{_{\hskip 0.01in}\bullet\bullet}}\over{#1}}
\def\ddt#1{\DDot{#1}}

\def\DDDot#1{\buildrel{_{_{\hskip 0.01in}\bullet\bullet\bullet}}\over{#1}}
\def\dddt#1{\DDDot{#1}}

\def\DDDDot#1{\buildrel{_{_{\hskip 
0.01in}\bullet\bullet\bullet\bullet}}\over{#1}}
\def\ddddt#1{\DDDDot{#1}}

\def\Tilde#1{{\widetilde{#1}}\hskip 0.015in}
\def\Hat#1{\widehat{#1}}

% Aligned equations

\newskip\humongous \humongous=0pt plus 1000pt minus 1000pt
\def\caja{\mathsurround=0pt}
\def\eqalign#1{\,\vcenter{\openup2\jot \caja
        \ialign{\strut \hfil$\displaystyle{##}$&$
        \displaystyle{{}##}$\hfil\crcr#1\crcr}}\,}
\newif\ifdtup
\def\panorama{\global\dtuptrue \openup2\jot \caja
        \everycr{\noalign{\ifdtup \global\dtupfalse
        \vskip-\lineskiplimit \vskip\normallineskiplimit
        \else \penalty\interdisplaylinepenalty \fi}}}
\def\li#1{\panorama \tabskip=\humongous                         % eqalignno
        \halign to\displaywidth{\hfil$\displaystyle{##}$
        \tabskip=0pt&$\displaystyle{{}##}$\hfil
        \tabskip=\humongous&\llap{$##$}\tabskip=0pt
        \crcr#1\crcr}}
\def\eqalignnotwo#1{\panorama \tabskip=\humongous
        \halign to\displaywidth{\hfil$\displaystyle{##}$
        \tabskip=0pt&$\displaystyle{{}##}$
        \tabskip=0pt&$\displaystyle{{}##}$\hfil
        \tabskip=\humongous&\llap{$##$}\tabskip=0pt
        \crcr#1\crcr}}

% Some defs

\def\eV{\,{\rm eV}}
\def\keV{\,{\rm keV}}
\def\MeV{\,{\rm MeV}}
\def\GeV{\,{\rm GeV}}
\def\TeV{\,{\rm TeV}}
\def\sv{\left<\sigma v\right>}
\def\({\left(}
\def\){\right)}
\def\cm{{\,\rm cm}}
\def\K{{\,\rm K}}
\def\kpc{{\,\rm kpc}}
\def\beq{\begin{equation}}
\def\eeq{\end{equation}}
\def\bea{\begin{eqnarray}}
\def\eea{\end{eqnarray}}

% New commands

\newcommand{\be}{\begin{equation}}
\newcommand{\ee}{\end{equation}}
\newcommand{\nbe}{\begin{equation*}}
\newcommand{\nee}{\end{equation*}}

\newcommand{\fr}{\frac}
\newcommand{\lb}{\label}

\thispagestyle{empty}

{\hbox to\hsize{
\vbox{\noindent March 2024 \hfill IPMU24-0005 \\
\noindent  \hfill }}

\noindent
\vskip2.0cm
\begin{center}

{\large\bf Dilaton-axion modular inflation in supergravity}

\vglue.3in

Daniel Frolovsky~${}^{a,*}$ and Sergei V. Ketov~${}^{a,b,c,\#}$ 
\vglue.3in

${}^a$~Interdisciplinary Research Laboratory, 
Tomsk State University, Tomsk 634050, Russia\\
${}^b$~Department of Physics, Tokyo Metropolitan University, Hachioji, Tokyo 192-0397, Japan \\
${}^c$~Kavli Institute for the Physics and Mathematics of the Universe (WPI),
\\The University of Tokyo Institutes for Advanced Study,  Chiba 277-8583, Japan\\
\vglue.2in

${}^{*}$~frolovsky@mail.tsu.ru, ${}^{\#}$~ketov@tmu.ac.jp
\end{center}

\vglue.3in

\begin{center}
{\Large\bf Abstract}  
\end{center}
\vglue.2in

\noindent Dilaton and axion are ubiquitous in extended supergravities and closed superstrings. We propose
new models of modular inflation in four-dimensional $N=1$ supergravity coupled to the chiral dilaton-axion-goldstino supermultiplet, which fit some necessary conditions of superstring cosmology. The model parameters are tuned to obey precision measurements of the cosmic microwave background radiation. We employ the modular-invariant superpotentials and asymptotically modular-invariant K\"ahler potentials, and achieve axion stabilization with high-scale supersymmetry breaking.

\newpage

\section{Introduction}

Cosmological inflation provides a great window into high-energy physics beyond the Standard Model (BSM) of elementary particles. However, this window is small at present, being limited to observations of the cosmic microwave background (CMB) radiation. Those observations are essentially confined to precision measurements of three observables: the amplitude of scalar perturbations $A_s$, the tilt $n_s$ of scalar perturbations, and the tensor-to-scalar-ratio $r$.~\footnote{The running $\a_s$ of $n_s$ is also an
observable whose observational bound is satisfied in our setup.}

The expected duration of inflation for about 60 e-folds implies a slow-roll of inflaton, which, in turn, leads to the existence of
a plateau in the potential and a related approximate shift symmetry  of the inflaton. It raises a question about the origin of the shift symmetry. The shift symmetry can be embedded into a larger symmetry group associated with the isometries of a non-linear sigma-model (NLSM) with a homogeneous target space, leading to multi-field inflation, which extends the space of inflation models even further. However, the inflaton shift symmetry may also have the environmental (not fundamental) origin. In any case, one needs a specific framework to address inflation together with BSM physics and, subsequently, minimize that framework  for more predictive power. Such framework exist and is given by supergravity. Supergravity can be considered as the low-energy limit of superstrings treated as quantum gravity. However, little is known about quantum gravity and strongly coupled superstrings where there is no computational control. Therefore, it is reasonable to confine ourselves to the supergravity framework by adding a few tools and general conditions inspired by superstrings and combine them with phenomenological conditions on the inflation model building. It is a 
non-trivial task in the context of axion-dilaton inflation models, whose structure is highly constrained.

In this Letter, we address those issues in the minimal framework of two-field (axion-dilaton) models of inflation in the four-dimensional $N=1$ supergravity. Our specific setup is given in Section 2. The new models are formulated in Section 3. Our main results are given in Section 4. Our conclusion is Section 5. We reduced the amount of references to a minimum because there are many related papers on the subject, see e.g., 
the reviews \cite{Lyth:1998xn,Baumann:2014nda,Cicoli:2023opf}  and references therein.

\section{The setup}

The scalar sector of a generic action for a single $N=1$ chiral matter multiplet minimally coupled to $N=1$ Einstein supergravity in four spacetime dimensions (after all fermions ignored and auxiliary fields eliminated) has the Lagrangian \cite{Wess:1992cp}~\footnote{We use the notation with the reduced Planck mass $M_{\rm Pl}=1$ and the spacetime signature $(-,+,+,+)$.}
\begin{equation}\label{lag1}
	\fracmm{\mathcal{L}_{\rm scalar}}{\sqrt{{-g}}}=-K_{,T\overline T}\partial^{\mu}T\partial_{\mu}\overline T-V(T,\overline T),
\end{equation}
where $T$ is a complex physical scalar, the K\"ahler metric $K_{,T\overline T}$ is given by
\begin{equation}
	K_{,T\overline T}=\fracmm{\partial^2K}{\partial T \partial \overline T}
\end{equation}
in terms of a K\"ahler potential $K(T,\bar{T})$, while the scalar potential reads \cite{Ferrara:1976ni}
\begin{equation}\label{vscal1}
V=e^K \left( K^{T\bar{T}} D_TW D_{\bar{T}}\Bar{W} -3\abs{W}^2\right)~,
\end{equation}
in terms of a superpotential $W(T)$, where we have used the definitions
\begin{equation} \label{not}
 K^{T\bar{T}} = (K_{,T \bar{T}})^{-1},\quad D_TW = \fracmm{\partial W}{\partial T}+W \fracmm{\partial K}{\partial T}~, \quad
 D_{\bar T}\Bar{W}= \fracmm{\partial\Bar{W}}{\partial{\bar T}}+\Bar{W} \fracmm{\partial K}{\partial{\bar T}}~.
\end{equation}

Unlike a generic (non-supersymmetric) Lagrangian for two scalars minimally coupled to Einstein gravity, the Lagrangian (\ref{lag1}) is significantly constrained by K\"ahler geometry in the field space. The input is just given by two potentials 
$K(T,\bar{T})$ and $W(T)$.

The K\"ahler potential 
\be \label{nosK}
K =  -3\a \ln \left(T +\bar{T}\right)
\ee
with the real parameter $\a>0$  leads to the kinetic term of a K\"ahler NLSM with
\begin{equation}\label{lkin1}
		 \fracmm{\mathcal{L}_{\rm kin.}}{\sqrt{{-g}}}=-\fracmm{3\alpha}{(T+\overline T)^2}\partial^{\mu}T\partial_{\mu} \overline T~,
\end{equation}
while it is known to arise in several different contexts in the literature. 

First, the $N$-extended pure supergravities in four spacetime dimensions have a scalar sector when $N\geq 4$, while the scalar sector of the $N=4$ supergravity \cite{Cremmer:1977tt} has the K\"ahler potential (\ref{nosK}). In other words, Eq.~(\ref{nosK}) can be considered as a consequence of extended local supersymmetry. 

Second, in closed superstring theories defined in ten spacetime dimensions, spacetime metric is always accompanied by a massless antisymmetric tensor (called B-field) and a massless scalar (called dilaton). The low-energy effective field theory of closed superstring theories is given by ten-dimensional supergravity. After compactification down to four spacetime dimensions, the B-field becomes axion that complexifies  dilaton to the $T$-field with the K\"ahler potential  (\ref{nosK}) also \cite{Becker:2006dvp}. 
The specific structure of Eq.~(\ref{nosK}) is often called "no-scale" in the literature, and  has the property
\be \lb{nospro}
K^{T\bar{T}} K_{,T}K_{,\bar{T}}=3\a=const.
\ee
In a compactified superstring theory, the value of $\alpha$ depends upon details of compactification of hidden dimensions.
For example, toroidal compactifications usually imply $3\alpha=3$ or $1$.

The no-scale K\"ahler structure was extensively used as the supergravity framework for sub-Planckian physics \cite{Ellis:2013nka}. In the context of superstring compactification, dilaton is related to the compactified volume of extra dimensions, so that its infinite value should correspond to the infinite volume or decompactification. The vacuum expectation value of dilaton determines the string coupling constant, whereas the vacuum expectation value of axion determines the amount of $CP$-violation.

Third, the higher-derivative supergravity, defined by local supersymmetrization of modified $F(R)$ gravity, can be transformed to the Einstein supergravity coupled to two chiral matter superfields (the inflaton superfield $T$ and the goldstino superfield $S$), while the $T$-dependence of the K\"ahler potential has the form (\ref{nosK}) too \cite{Cecotti:1987sa,Ketov:2023ykf}. The no-scale structure of the K\"ahler potential is particularly useful for describing inflation in supergravity  because it effectively solves the so-called $\eta$-problem by eliminating the exponential factor $e^K$ present in Eq.~(\ref{vscal1}) in the case of the canonical K\"ahler potential $K=\Bar{T} T$. It is possible to identify the inflaton superfield with the goldstino superfield thus reducing the number of physical degrees of freedom \cite{Ketov:2014qha}. We use the minimal setup where both inflaton/dilaton, axion and goldstino are in the same chiral supermultiplet $T$, which has only one complex scalar.

Having the K\"ahler potential (\ref{nosK}) with a superpotential $W$, we find from Eq.~(\ref{vscal1}) the potential
\begin{equation} \label{pot}
	V=(3\alpha-1)\fracmm{W \overline W}{(T+\overline T)^{3\alpha}}+\fracmm{W_{,T}\overline W_{,\overline T}}{3\alpha(T+\overline T)^{3\alpha-2}}-\fracmm{\overline W W_{,T}+W\overline W_{, \overline T}}{(T+\overline T)^{3\alpha-1}}~~.
\end{equation}

In supergravity, both K\"ahler potential and superpotential can be modified and receive quantum corrections as well, though the superpotential does not get loop corrections due to the non-renormalization theorem in supersymmetry
\cite{Wess:1992cp}.  Deriving the exact potentials $K$ and $W$ from superstring theory remains one of the major problems preventing reliable phenomenological predictions. The potential of the dilaton-axion $T$ vanishes in perturbative superstring theory, whereas its nonperturbative derivation is beyond computational control, see e.g., Refs.~\cite{Dine:1985he,Alexandrov:2016plh}.  Because of that, one usually comes with certain assumptions about the structure of $K$ and $W$.

We assume (i) the K\"ahler potential has the no-scale structure (\ref{nosK}) for very small and very large dilaton values, being modified for finite dilaton values describing inflation, (ii) the superpotential is modular-invariant, and (iii) the potential should vanish for large dilaton values. The conditions (i) and (iii) are motivated for large dilaton values because the asymptotic potential is under control in weakly coupled superstring theory, which implies the modular invariance and the vanishing potential. As is also known in superstring theory, the non-perturbative dilaton potential must have the run-away behavior \cite{Dine:1985he}, so that one needs a significant potential barrier in order to separate a phenomenologically viable region at strong coupling for finite values of dilaton from the spacetime decompactification region at weak coupling for large dilaton values. The  assumption (ii) does not follow from superstring theory, being motivated by an idea that the modular invariance is relevant for the superpotential also, {\it cf.} Ref.~\cite{Schimmrigk:2016bde,Schimmrigk:2021tlv,Abe:2023ylh}.
 Our goal is to investigate whether a realistic potential in the no-scale supergravity under those restrictive conditions can be found, which would be suitable for describing inflation in the early Universe and dark energy (as a de Sitter vacuum) in the current Universe.
 
The NLSM (\ref{lkin1}) has the field target space isomorphic to the coset $SL(2;\mathbb R)/SO(2)$ or 
$SU(1,1; \mathbb C)/U(1)$ with a constant negative curvature (also known as Lobachevsky plane with hyperbolic geometry),
and, hence, it has the continuous symmetry (or isometry) $SL(2,\mathbb R)$.\, see e.g., Ref.~\cite{Ketov:2000dy} for more.
The action of this symmetry on the $T$-field is given by
\begin{equation}\label{sym}
	T \rightarrow \fracmm{aT+ ib}{icT+d}~~,
	\qquad \quad 
	\begin{array}{cc}
		 a,b,c,d \in \mathbb R~~, \\
		 ad+bc=1~~,
	\end{array}
\end{equation}
that includes
\begin{eqnarray*}
		 \text{imaginary translations:} \qquad   &T\rightarrow T+ib~,
		\qquad  &a=d=1, \quad c=0~, \\
	 \text{dilatations:} \qquad  &T\rightarrow a^2 T, \qquad  &c=b=0, \quad d=a^{-1}~, \\
	  \text{inversions:} \qquad   &T\rightarrow \left( \fracmm{b}{c}\right)\fracmm{1}{T}~, \qquad  &a=d=0, \quad bc=1~. 	
\end{eqnarray*}

 To make a connection to the mathematical literature, let us redefine 
 \begin{equation}
 	T=-i S\equiv -i\tau
 \end{equation}
 and $a=\alpha$,  $b=-\beta$, $c=\gamma$, $d=\delta$. Then Eq.~(\ref{sym}) takes the standard form
 \begin{equation}\label{symm}
	S \rightarrow \fracmm{\alpha S+ \beta}{\gamma S+\delta}~~,
	\qquad \quad 
	\begin{array}{cc}
		 \alpha,\beta,\gamma,\delta \in \mathbb R~, \\
		 \alpha\delta-\beta\gamma =1~~,
	\end{array}
\end{equation}
while the kinetic term (\ref{lkin1}) in terms of $S$ reads
\begin{equation}\label{lkin2}
			\fracmm{\mathcal{L}_{\rm kin.}}{\sqrt{{-g}}}
			=\fracmm{3\alpha}{(S-\overline S)^2}\partial^{\mu}S\partial_{\mu} \overline S~.
\end{equation}
 The dilaton field $\varphi$ and the axion field $\sigma$  can be introduced as follows:
 \begin{equation}\label{tau}
 	S(x)=\sigma(x)+ie^{\sqrt{\frac{2}{3\alpha}}\varphi(x)} \equiv \tau(x)~,
 \end{equation}
which leads to the kinetic terms
\begin{equation}
\fracmm{\mathcal{L}_{\rm kin.}}{\sqrt{{-g}}}	
	=-\fracmm{1}{2}\partial^{\mu}\varphi\partial_{\mu}\varphi-\fracmm{3\alpha}{4} e^{-2\sqrt{\frac{2}{3\alpha}}\varphi}
	\partial^{\mu}\sigma\partial_{\mu}\sigma
\end{equation}
having the canonical form for $\varphi$. Translations in Eq.~(\ref{sym}) become a shift symmetry,
 \begin{equation}
 	\sigma(x)\rightarrow \sigma(x)+\beta~
 \end{equation}
 so that $\sigma(x)$ is an axion. Dilatations in Eq.~(\ref{sym}) imply that $\varphi$ is dilaton.
 
Of course, the continuous symmetry (\ref{sym}) is incompatible with any potential except a constant. Therefore, it must be broken. However, it may be expected that its {\it discrete} subgroup $SL(2,\mathbb Z)$ survives in superstring theory based on superconformal quantum field theory in two dimensions \cite{Becker:2006dvp,Ketov:1995yd}. The invariance under the $SL(2,\mathbb Z)$ is known as the modular invariance. It is generated by two transformations:
\be \label{tdual}
\tau \to \tau +1~,
\ee
known as T-duality, and 
 \begin{equation}\label{sdual}
 	\tau \rightarrow - \fracmm{1}{\tau}~,
 \end{equation}
  known as S-duality.  
 
 The K\"ahler metric (not the K\"ahler potential!) in Eq.~(\ref{lkin2}) is also invariant under the $SL(2,\mathbb Z)$ transformations, while it is also possible to have a non-trivial (meromorphic) superpotential invariant under the $SL(2,\mathbb Z)$ by using the
 standard (Eisenstein) series or the modular forms defined in the upper (complex) half-plane $\tau$,
 \begin{equation}
 	E_4(\tau)=1+240\sum_{n=1}^{\infty}\fracmm{n^3 q^n}{1-q^n}, \qquad E_6(\tau)=1-504\sum_{n=1}^{\infty}\fracmm{n^5 q^n}{1-q^n}, \qquad q=e^{2\pi i \tau}.
 \end{equation}
 They are the special modular forms $F_k(\tau)$ of weight $k$, which transform under the $SL(2,\mathbb Z)$ according to the rule
\begin{equation}
	F_k\left(\fracmm{\alpha \tau+ \beta}{\gamma \tau+\delta}\right)=(\gamma \tau+\delta)^kF_d(\tau).
\end{equation}
In particular, one has
\begin{equation}
	F_k(\tau+1)=F_k(\tau) \quad {\rm and} \quad F_k\left(\fracmm{1}{-\tau}\right)=\tau^kF_k(\tau).
\end{equation}
 
 It is remarkable that there exists only one modular function (of zero weight $k=0$) that is holomorphic everywhere on the upper-half plane expect the point at $i\infty$, where is has a pole.  This  function is known in the mathematical literature as the (Klein) $j$-invariant   
  \begin{equation}
 	j(\tau)=1728\fracmm{E_4(\tau)^3}{E_4(\tau)^3-E_6(\tau)^2}~~.
 \end{equation}
 It can be expanded into Laurent series near the pole,
 \begin{equation}
 	j(\tau)=q^{-1}+744+196884 q+21493760 q^2+864299970 q^3+20245856256 q^4+\ldots 
 \end{equation}

 Any analytic function of $j(\tau)$ also has similar properties, so that we should consider the superpotentials $W$ on the space of all regular functions of $j(\tau)$ or its normalized cousin $J(\tau)=j(\tau)/1828$, i.e. $W(T)= f(J(iT))$, and then impose physical (or phenomenological) conditions on our choice of the function $f(J)$. This approach is known as the  modular inflation in the literature, see e.g., Ref.~\cite{Schimmrigk:2021tlv} and the references therein. Unlike  Ref.~\cite{Schimmrigk:2021tlv} with
 $V\sim \abs{j(\tau)}^2$ , we use the supergravity framework with its holomorphic structure.

 \section{The models}
 
 The phenomenological conditions we impose on the scalar potential are as follows:
\begin{itemize}
\item the potential should have a plateau of positive height and a Minkowski vacuum for finite dilaton values,
\item the potential should have the run-away behaviour and approach zero for very large dilaton values,
\item axion should be stabilized during inflation,
\item the potential should realize viable inflation consistent with observations of the cosmic microwave background radiation,
\item the inflation plateau should be separated from the run-away part by a high potential barrier of finite height,
\item supersymmetry should be spontaneously broken.
\end{itemize}
  
It is worth mentioning that obeying all those conditions is by no means guaranteed (it is hard actually) in the supergravity setup of Sec.~2 because the only tools at our disposal are (i) a choice of the $f(J)$ function in the superpotential, and (ii) a modification
of the K\"ahler potential (\ref{nosK}) without loosing the asymptotic modular invariance of the NLSM metric.

To begin with, let us consider the asymptotic behavior  of the scalar potential (\ref{pot}) with $W\propto J$ when $\tau \rightarrow i\infty$ along a possible inflationary trajectory with $T=\overline T=t$. We have $J(it)\rightarrow e^{2\pi t}$ for $t\rightarrow +\infty$,
so that Eq.~(\ref{tau}) implies the rapidly increasing double exponential behavior of the potential of the canonical inflaton 
(dilaton) $\varphi$, which is obviously too steep. Simultaneously, this gives us a clue that the $f(J)$ function should be constructed out of the logarithms of $J$,  perhaps, being multiplied by other analytic functions with limited values, 
like powers of $\tanh(J)$, in the superpotential. We adopt the strategy to find such functions, obeying the above phenomenological conditions,  as a combination of analytical proposals with machine learning to fix the shape of those functions. 
  
 When considering a simple ansatz for the superpotential in the form
 \begin{equation}\label{sW}
 	W(T)=M\ln^m[J(i\cdot T)]
 \end{equation}
 with the two parameters $M$ and $m$, we find
  \begin{equation}
 	 \ln^m{[J(it)]}_{t \rightarrow +\infty}\simeq  t^m 
 \end{equation}
 and
 \begin{equation} \label{aspot}
 	V_{t \rightarrow +\infty}\simeq \fracmm{1}{2^{3\alpha-2}}\left(\fracmm{3\alpha-3}{4}+\fracmm{m^2}{3\alpha}-m\right)t^{2m-3\alpha}.
 \end{equation}
 Therefore, there are two possibilities to get $V\rightarrow 0$ for $t\rightarrow+\infty$: either (a) with $2m<3\alpha$ or (b)
 with the vanishing coefficient in the brackets, which leads to $m=\fracmm{3}{2}(\alpha\pm \sqrt{\a})$. We found the case (b) too restrictive because it did not allow us to reach all the goals formulated at the beginning of this Section, even  with a modified K\"ahler potential.~\footnote{The case (b) with the modified K\"ahler potential --- see Eq.~(\ref{modK}) below --- and a power-like superpotential $W\propto T^m$ can describe viable inflation in supergravity \cite{Pallis:2023aom} but it does not satisfy the other requirements given at the beginning of this Section and motivated by superstring theory.}
 
\begin{figure}[h]
\begin{minipage}[h]{0.46\linewidth}
\center{\includegraphics[width=1\textwidth]{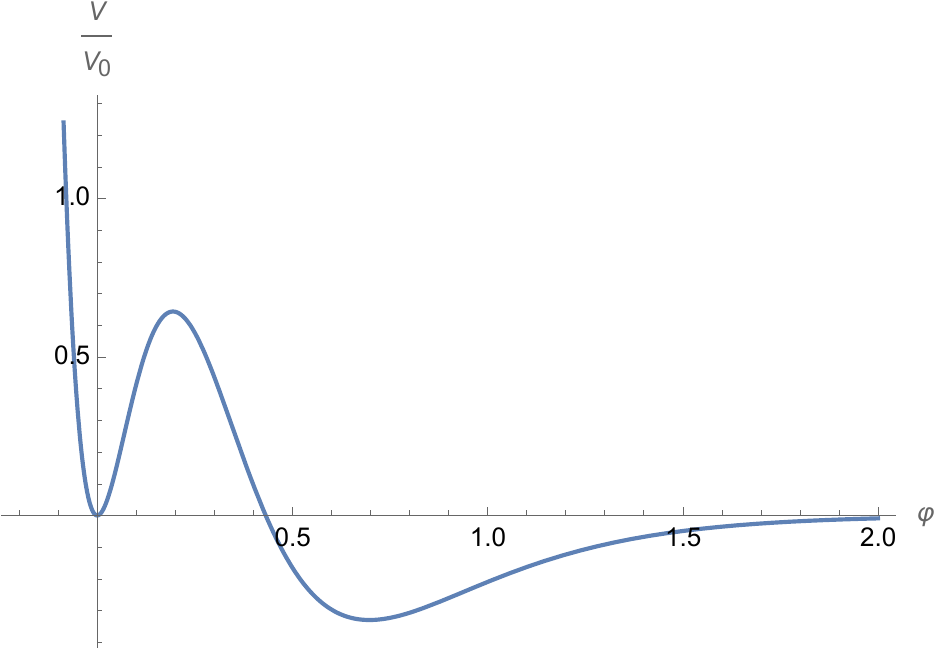}}
\end{minipage}
\hfill
\begin{minipage}[h]{0.46\linewidth}
\center{\includegraphics[width=1\textwidth]{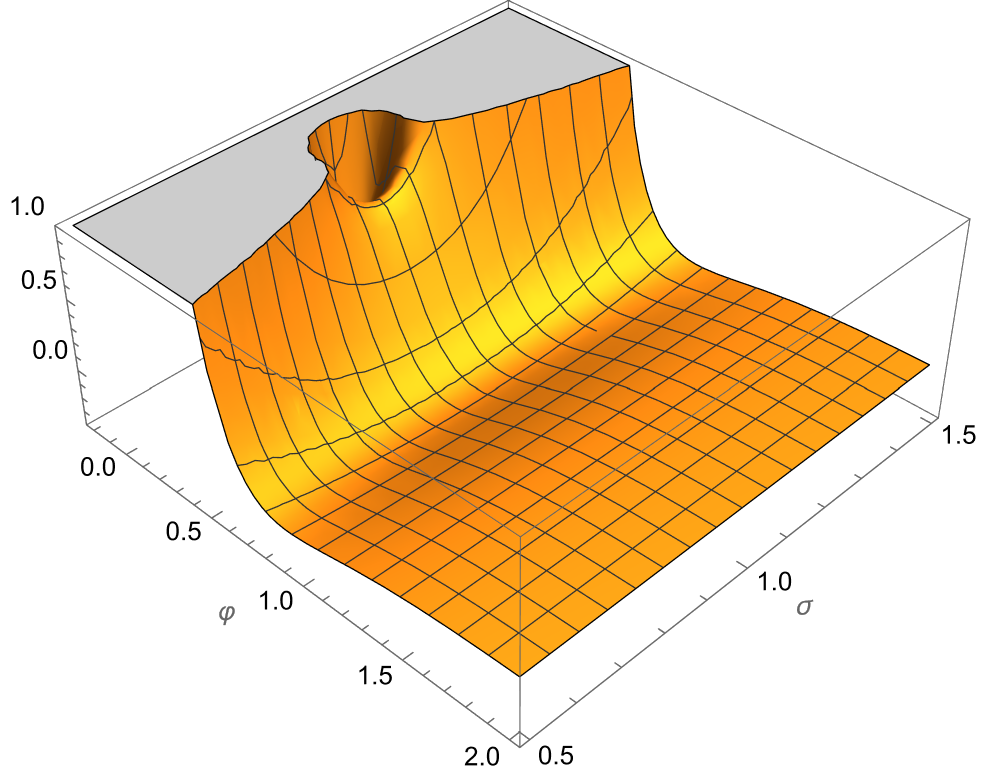}}
\end{minipage}
\caption{The left figure shows a slice of the scalar potential $V$ with the superpotential (\ref{sW}) along the dilaton axis $\varphi$ with an integer value of axion $\sigma$. The right figure shows the three-dimensional plot of the potential $V$ with $\alpha =1$, $m=1$ and $V_0\sim M^2$.}
\label{fig1}
\end{figure}

A typical  example of the potential in the case (a) is illustrated by Fig.~\ref{fig1}. It has the anti-de-Sitter (AdS) vacuum that is  typical in supergravity, as may have been expected. Its uplifting to a dS vacuum was usually achieved in the literature by invoking other interactions \cite{Kachru:2003aw}. Alternatively, we can avoid an AdS vacuum for the relevant range of dilaton values in the scalar potential by setting the parameter $\alpha$ to a sufficiently low value (see the top lines in Fig.~\ref{fig2}). However, lowering the parameter $\alpha$ is limited from below because it should not violate the condition (a): $2m<3\alpha$.

 \begin{figure}[h]
\begin{minipage}[h]{0.46\linewidth}
\center{\includegraphics[width=1\textwidth]{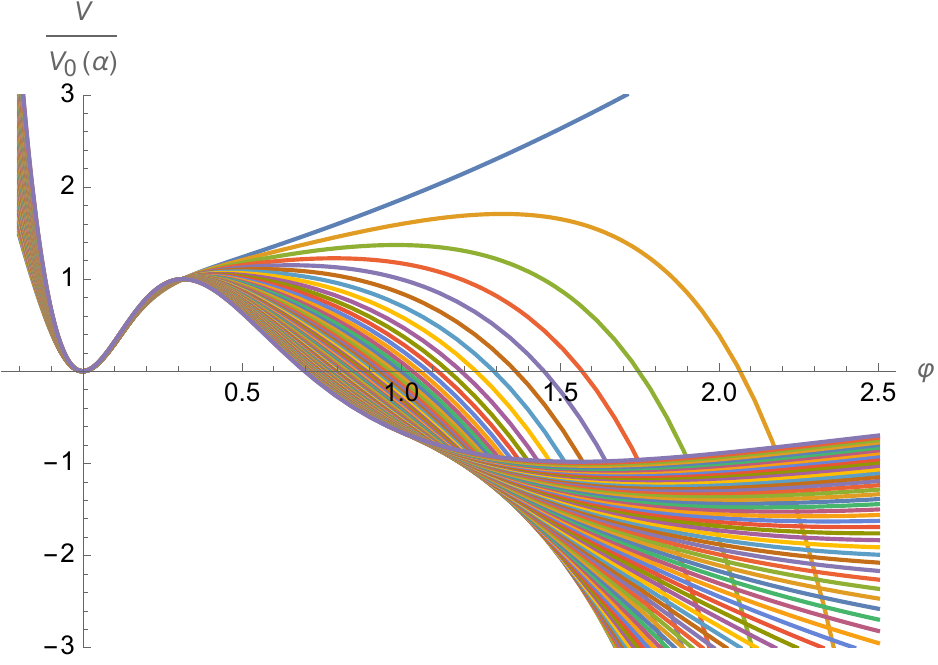}}
\end{minipage}
\hfill
\begin{minipage}[h]{0.46\linewidth}
\center{\includegraphics[width=1\textwidth]{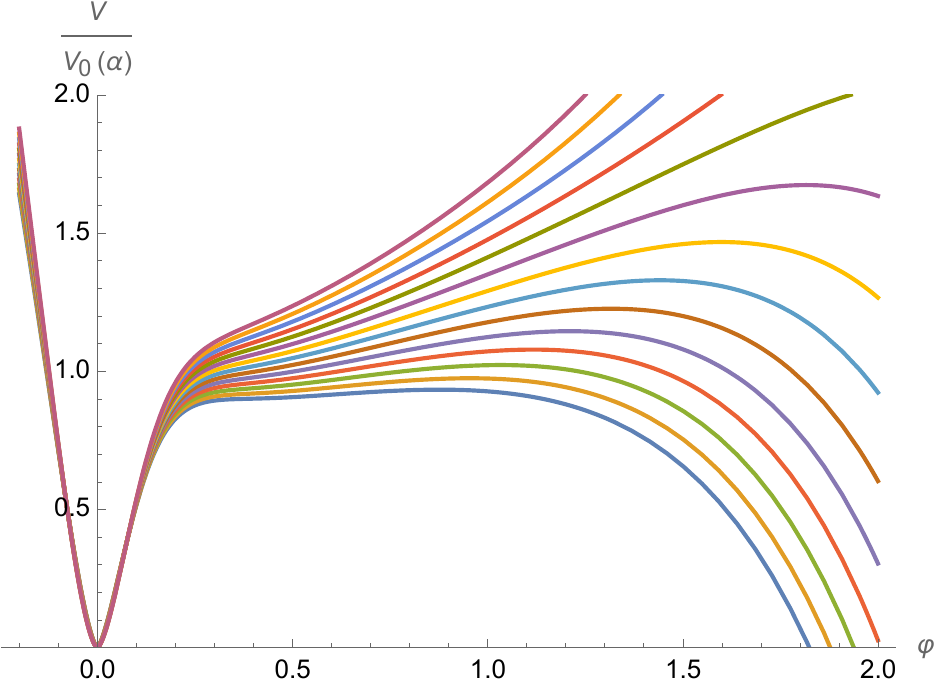}}
\end{minipage}
\caption{The left figure shows the dependence of $V/V_0$ upon $\alpha$ in the range $\a\in [1,\ldots, 0.1]$ with $m=1$. The right figure shows the potential dependence upon $m$ in the range $m\in [0.9,\ldots , 0.86]$ with $\alpha=0.175$. Both figures correspond to the model with the K\"ahler potential (\ref{nosK}) and the superpotential (\ref{sW}).}
\label{fig2}
\end{figure}

When exploring the space of parameters $(\a,m)$ satisfying the condition $2m<3\alpha$, we find altering these parameters does not eliminate an Anti-de Sitter (AdS) vacuum. Instead, it shifts the vacuum towards larger dilaton values when attempting to create a plateau for finite values of dilaton. Hence, to make the potential more realistic, one may try to hide that AdS vacuum behind a potential barrier. However, this is only possible by modifying  the K\"ahler potential. For those purposes, we employ the modified K\"ahler potential proposed in Ref. \cite{Pallis:2023aom} and having the form
\begin{equation}\label{modK}
		K=-3 \alpha\ln\left[ T+\overline T+\xi^2(T+\overline T-2\nu )^4\right]~,
\end{equation}
where the extra term was added under the logarithm with two new parameters $(\x,\n)$. Besides leading to a plateau in the scalar potential and stabilizing the axion during inflation along the trajectory  $T=\Bar{T}$, the modified K\"ahler potential still has the asymptotic modular invariance for small ${\rm Re}\,T$ as well as for large ${\rm Re}\,T$ though with $\a \to 4\a$.~\footnote{The K\"ahler potential (\ref{modK}) does not break the axion-shift symmetry, being different from the K\"ahler potentials proposed  in Ref.~\cite{Carrasco:2015uma}.} It is also worth mentioning that an AdS vacuum behind the potential barrier always implies the asymptotic potential approaching zero at infinity from below, not from above, leaving only one
option out of two mentioned in Ref.~\cite{Dine:1985he}.

Given the modified K\"ahler potential  (\ref{modK}), the parametrization (\ref{tau}) is no longer valid to get the canonical inflaton $\varphi$. When using the ansatz $\rm Re\,T = F(\varphi)$  with some function $F(\varphi)$, one has to solve the differential equation
\begin{equation}\label{par}
	\fracmm{3 \alpha  \left(1024 \xi ^4 (\nu-2 F(\varphi))^6-128 \xi ^2 (2 F(\varphi )+2 \nu) (\nu-2 F(\varphi ))^2+4\right)}{4 \left(8 \xi ^2 (\nu-2 F(\varphi ))^4+2 F(\varphi )\right)^2}\left(\fracmm{\partial F(\varphi)}{\partial \varphi}\right)^2=\fracmm{1}{2}~~.\end{equation}
We solved this equation numerically, with the result illustrated by Fig.~\ref{fig3} that shows a difference between 
$\rm Re\,T$ from Eq.~(\ref{tau}) and $F[\varphi]$.
	
\begin{figure}[h]
\begin{center}
  \includegraphics[width=0.46\textwidth]{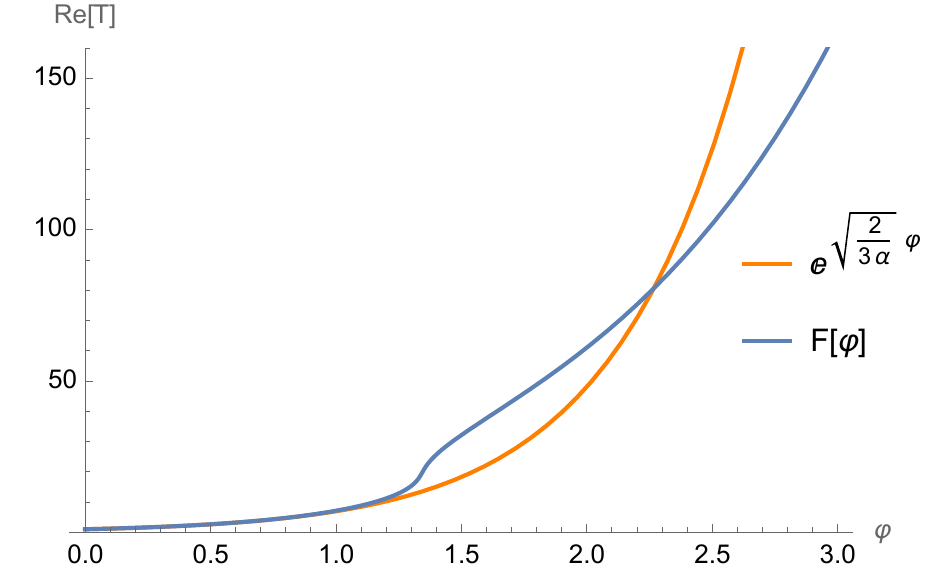}
  \caption{The blue line shows a numerical solution to Eq.~(\ref{par}) needed for the canonical parametrization of dilaton/inflaton field. The red line shows the parametrization of the dilaton field used in Eq.~(\ref{tau}). The parameters take the values 
  $\alpha=0.1753$, $\nu=-1$ and $\xi=0.001$. }
   \label{fig3}
  \end{center}
\end{figure}

Spontaneous supersymmetry  breaking occurs when the vacuum expectation value of the complex auxiliary field $F_T$ 
in the off-shell chiral supermultiplet $T$ does not vanish, $\VEV{F_T}\neq 0$. Then gravitino acquires the mass $m_{3/2}=
\VEV{e^{G/2}}$  by eating up goldstino (it is known as super-Higgs effect \cite{Wess:1992cp} in the literature), 
where $G=K+\abs{W}^2$. In our models the scale of supersymmetry breaking is comparable to the scale of inflation,
with both being related to the parameter $M$.

\section{The results}

When using the modified K\"ahler potential (\ref{modK}) with the superpotential (\ref{sW}), we numerically obtained the potentials displayed in Fig.~\ref{fig4}. All of them have a plateau and a potential barrier separating the plateau from the run-away region. The false vacuum before the barrier can be either AdS, dS or Minkowski, or it can be absent at all.

\begin{figure}[h]
\begin{minipage}[h]{0.46\linewidth}
\center{\includegraphics[width=1.1\textwidth]{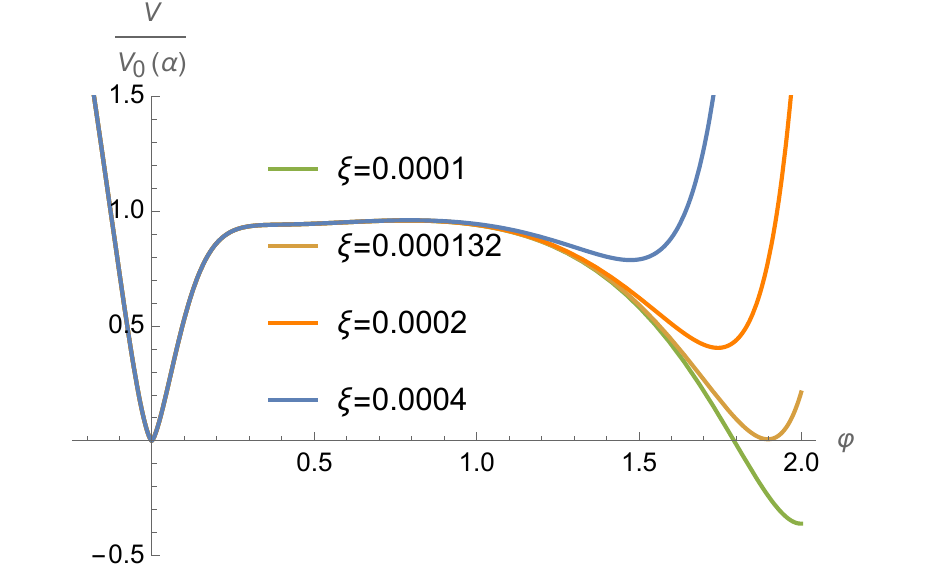}}
\end{minipage}
\hfill
\begin{minipage}[h]{0.46\linewidth}
\center{\includegraphics[width=1.1\textwidth]{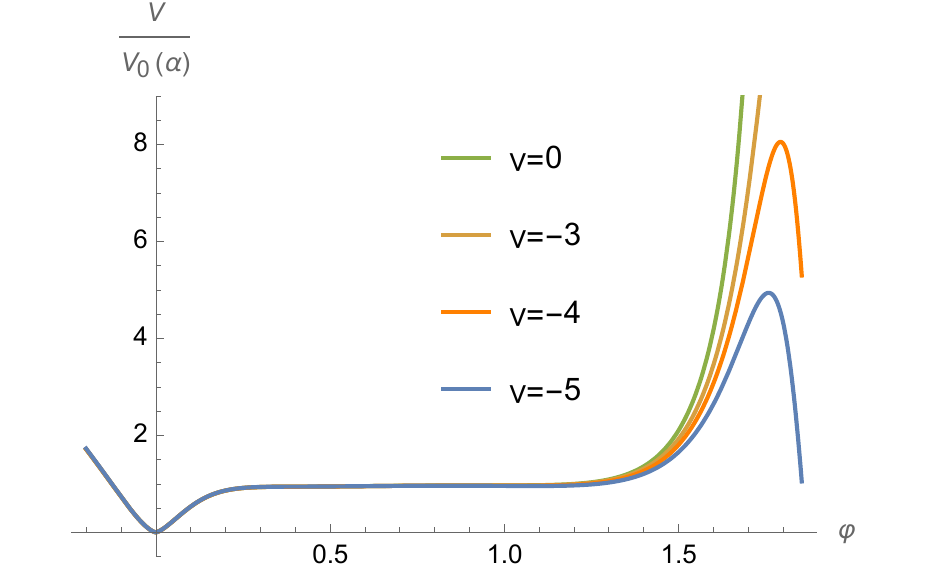}}
\end{minipage}
\caption{The left picture shows the dependence of $V/V_0$ upon $\xi$ that controls the position of the barrier and the false vacuum. The right figure shows the dependence of $V$ upon $\nu$, when the other parameters take the values $m=0.8637$, $\xi=0.001$ and $\alpha=0.1753$.}
\label{fig4}
\end{figure}

More general superpotentials $W(J)$ preserving the modular invariance are also possible and lead to additional features such as a local near-inflecton point  during inflation, which might lead to production of primordial black holes in supergravity \cite{Ketov:1995yd}. We found two new classes of them having the form
\be \label{twopot}
(i):~W=M\ln^m{(J(iT))}\tanh^{m_1}{(J(iT))}~,\quad    (ii):~W=M\ln^m{(n J(iT))}\ln{J(iT)}~,
\ee
with extra parameters $(m_1,n)$. The plots of some potentials are displayed in Fig.~\ref{fig5}.

\begin{figure}[h]
\begin{minipage}[h]{0.46\linewidth}
\center{\includegraphics[width=1.1\textwidth]{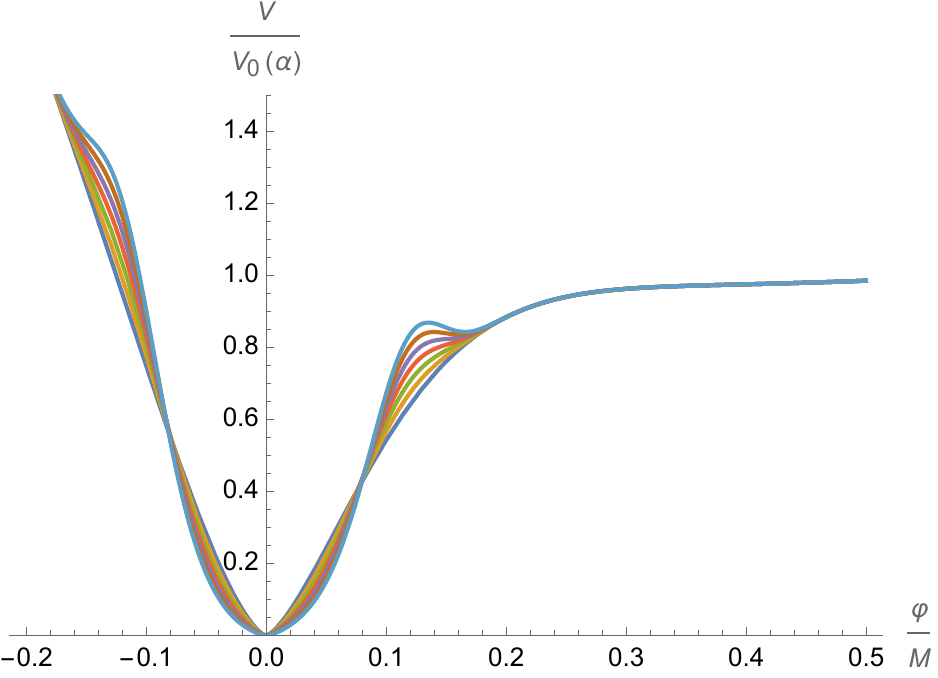}}
\end{minipage}
\hfill
\begin{minipage}[h]{0.46\linewidth}
\center{\includegraphics[width=1.1\textwidth]{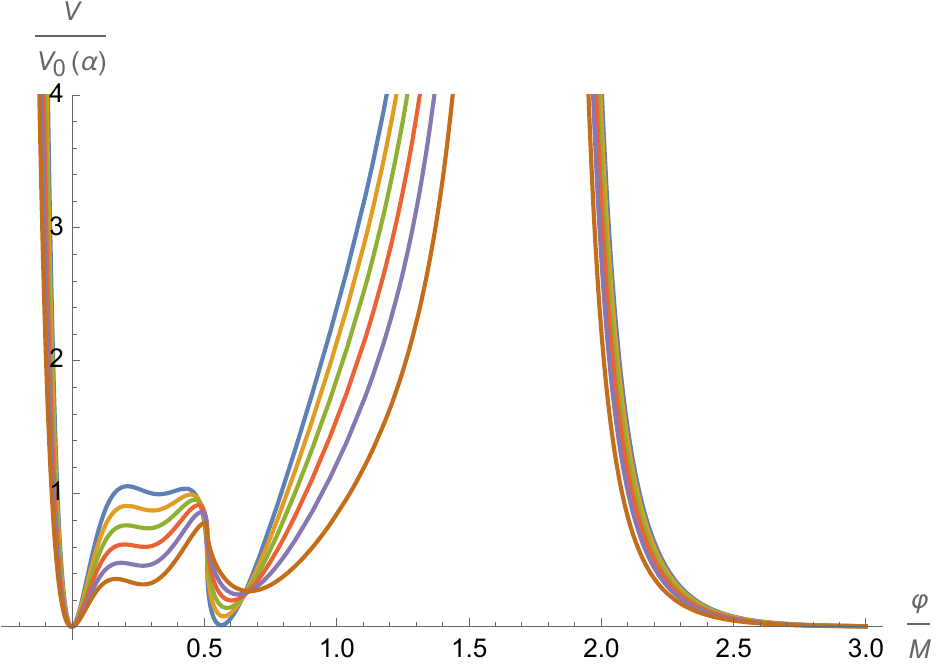}}
\end{minipage}
\caption{The left figure shows the dependence of $V/V_0$  upon $m_1$ in the case (i) with the range $m_1\in [0.5,\ldots, 2]$ that controls the  height of the bump. The right figure in the case (ii) shows the dependence of $V/V_0$ upon $n$ in the range $n\in [-1.5,\ldots ,-6.5]$  that controls the inflection of the potential.  The other parameters take the values  $m=0.8637$ and 
$\alpha=0.1753$ in the case (i), and $m=3$, $\alpha=4$ and $\xi=0.041$ in the case (ii), respectively.}
\label{fig5}
\end{figure}

In both cases, changing the parameters $\xi$ and $\nu$ affects the scalar potential in the same way shown in Fig.~\ref{fig3}. The $\log(J)$ factor predominantly determines the shape of the scalar potential on the inflation scale. The free parameter $M$ allows us to match any desired scale of inflation in agreement with the observed amplitude of scalar perturbations in the CMB. We found our derived potentials did not lead to a significant production of primordial black holes because of insignificant enhancement of the power spectrum of scalar perturbations.

Inflationary dynamics in our models with two scalars can be studied along the standard lines, see e.g., Section 5 of Ref.~\cite{Aldabergenov:2020bpt}, starting from the quintessence action
\begin{equation}
	e^{-1} \mathcal{L}=\frac{1}{2} R-\frac{1}{2} G_{A B} \partial \phi^A \partial \phi^B-V,
\end{equation}
with the K\"ahler NLSM metric $G_{AB}$ and the potential $V$, where  $\phi^A=\{\varphi, \sigma\}$ and $A=1,2$. The NLSM 
metric is given by
\begin{equation}
G_{A B}=\left(\begin{array}{cc}
1 & 0 \\
0 &  \fracmm{3 \alpha  \left(1024 \xi ^4 (\nu-2 F[\varphi])^6-128 \xi ^2 (2 F[\varphi ]+2 \nu) (\nu-2 F[\varphi ])^2+4\right)}{2 \left(8 \xi ^2 (\nu-2 F[\varphi ])^4+2 F[\varphi ]\right)^2}
\end{array}\right).
\end{equation}

The background equations of motion, the equations for perturbations, the standard definitions of the effective mass matrix, of the adiabatic and isocurvature parameters, of the transition functions, and of the Hubble flow parameters can be found  in Ref.~\cite{Aldabergenov:2020bpt}. When using the superpotential (\ref{sW}) and  the K\"ahler potential (\ref{modK}), we got the results displayed in Fig.~\ref{fig6} that shows the effective dynamics is essentially of the single-field type.

\begin{figure}[h]
\begin{minipage}[h]{0.46\linewidth}
\center{\includegraphics[width=1.1\textwidth]{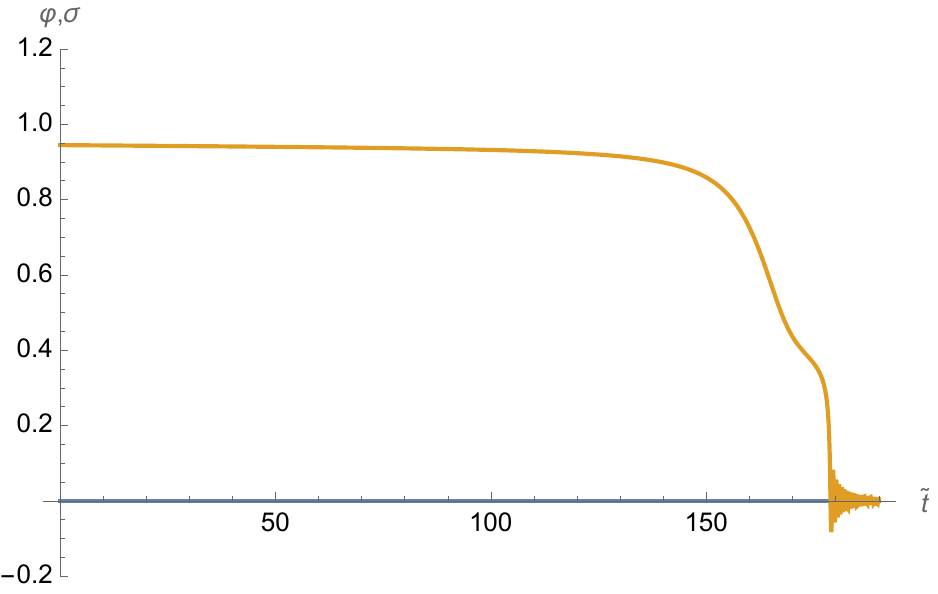}}
\end{minipage}
\hfill
\begin{minipage}[h]{0.46\linewidth}
\center{\includegraphics[width=1.1\textwidth]{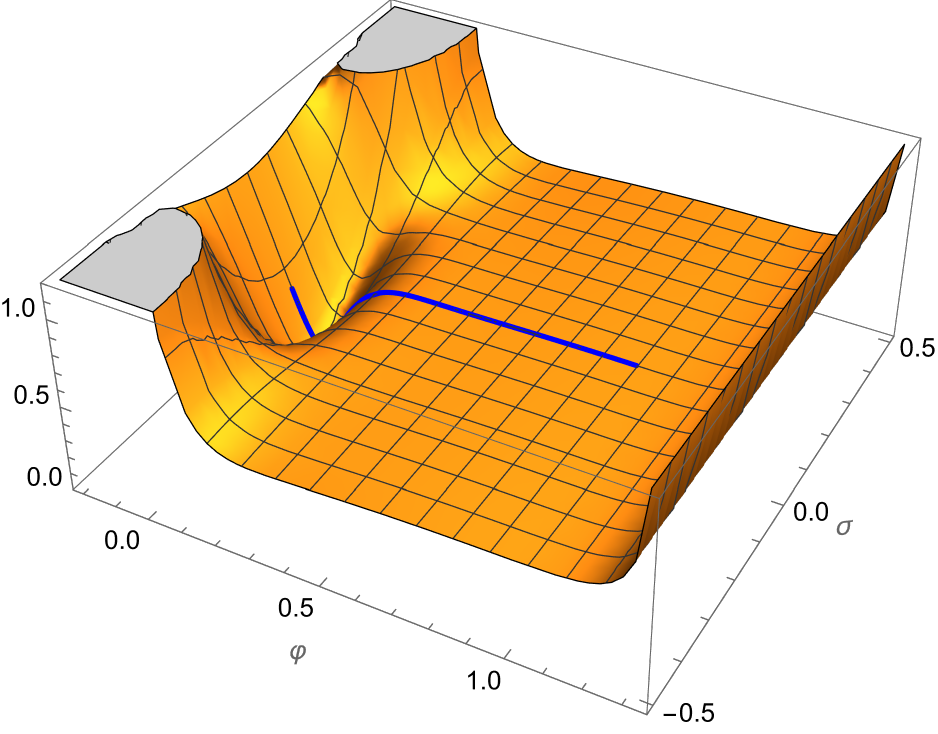}}
\end{minipage}
\caption{The left figure shows the evolution of dilaton and axion fields. The right figure shows the field trajectory on the surface of the scalar potential $V/V_0$. The parameter values are  $\alpha=0.1779758$, $m=0.863775$, $\xi=0.001$  and $\nu=-1.$ }
\label{fig6}
\end{figure}

The inflationary observables (CMB tilts) in our supergravity models were  numerically computed at  60 e-folds before the end of inflation by using the transport method \cite{Dias:2015rca} with the results 
\be \lb{nsr}
n_s=0.9649 \quad  {\rm and} \quad r\leq 4.1\cdot10^{-7}~,
\ee
for the tilt $n_s$ of scalar perturbations and  the tensor-to-scalar ratio $r$, respectively, which are fully consistent with the CMB measurements
\cite{BICEP:2021xfz,Tristram:2021tvh}:
\begin{equation}
	n_s= 0.9649 \pm0.0042 \quad (68\% \text{C.L.})~,  \quad r<0.032 \quad (95\% \text{C.L.})~,
\end{equation}
when using the parameters $\alpha=0.1779758$, $m=0.8637775$, $\xi=0.001$, $\nu=-1$, and the initial conditions 
$\varphi(0)=0.945$, $\sigma(0)=0$, $\dot\varphi(0)=\dot\sigma(0)=0$. The dependence upon the initial conditions was weak. The running index $\a_s$ of $n_s$ is of the order $10^{-3}$ in our models.
% similarly to Ref.~\cite{Pallis:2023aom}.

We used the machine learning algorithms to fine-tune the parameters and flatter the shape of the potential at the scale of inflation.

\section{Conclusion}

In this Letter we proposed new concrete models of modular inflation by using the minimal supergravity setup in the dilaton-axion scalar sector. We combined the phenomenological requirements with the limited input from superstring cosmology. Though our cosmological models were not derived from superstrings, our results in supergravity demonstrate the existence of realistic models of inflation in the highly constrained framework which was left as no-go in the earlier work in supergravity and string theory, see e.g., Ref.~\cite{Lyth:1998xn,Baumann:2014nda,Cicoli:2023opf} and the references therein.

An ultra-violet (UV) completion of our supergravity models in superstring theory is beyond the scope of this investigation. In our models, the mass parameter $M$ defining the inflation scale should be at least five orders of the magnitude lower than the Planck mass, whereas the UV-cutoff is expected to be given by the Planck mass. When assuming the quantum gravity (string) scale and the Kaluza-Klein scale to be much higher than the inflation scale,  quantum corrections to our supergravity models should not have a significant impact.

\section*{Acknowledgements}

One of the authors (SVK) acknowledges discussions with Yermek Aldabergenov and Costas Pallis.

This work was partially supported by Tomsk State University under the development program Priority-2030.  SVK was also supported by Tokyo Metropolitan University, the Japanese Society for Promotion of Science under the grant No.~22K03624, and the World Premier International Research Center Initiative, MEXT, Japan.

\bibliography{Bibliography}{}
\bibliographystyle{utphys}

\end{document}